\documentclass[11pt]{article}

\usepackage{url}
\usepackage{graphicx}
\usepackage{comment}
\usepackage{fullpage}

\begin{document}

\title{A Generic Checkpoint-Restart Mechanism for Virtual Machines}

\author{Rohan Garg and Komal Sodha and Gene Cooperman\\
        Northeastern University\\
        Boston, MA, USA\\
        \{rohgarg,komal,gene\}@ccs.neu.edu }
\date{}
% AUTHORS:  Rohan Garg and Komal Sodha and Gene Cooperman
% ACKNOWLEDGMENT:  Zhengping Jing

\maketitle

\begin{abstract}

It is common today to deploy complex software inside a virtual
machine~(VM).  Snapshots provide rapid deployment, migration between
hosts, dependability (fault tolerance), and security (insulating a guest
VM from the host).  Yet, for each virtual machine, the code for snapshots
is laboriously developed on a per-VM basis.  This work demonstrates a
generic checkpoint-restart mechanism for virtual machines.  The mechanism
is based on a plugin on top of an unmodified user-space checkpoint-restart
package, DMTCP.  Checkpoint-restart is demonstrated for three virtual
machines:  Lguest, user-space QEMU, and KVM/QEMU.  The plugins for Lguest
and KVM/QEMU require just 200~lines of code.  The Lguest kernel driver
API is augmented by 40~lines of code.  DMTCP checkpoints user-space
QEMU without any new code.  KVM/QEMU, user-space QEMU, and DMTCP need
no modification.  The design benefits from other DMTCP features and
plugins. Experiments demonstrate checkpoint and restart in 0.2~seconds
using forked checkpointing, mmap-based fast-restart, and incremental
Btrfs-based snapshots.

\end{abstract}

\section{Introduction}

A generic mechanism is presented for checkpointing virtual machines.
Snapshots of virtual machines are a key technology for dependable
computing.  They are more important today than ever for deployment in
clouds (including IaaS, ``Infrastructure as a Service''), rapid deployment
(starting from an initial snapshot), migration between hosts, fault
tolerance for dependability, and greater security by insulating a guest
VM from the host.  Current virtual machines rely on machine-specific
checkpoint-restart mechanisms.  Such designs struggle with common
checkpointing issues (live checkpointing without stopping the virtual
machine, incremental checkpointing, differential checkpointing, forked
checkpointing (checkpointing within a forked child process) concurrently
with execution within a parent, and checkpointing of distributed virtual
machines.

By employing a standard checkpoint-restart package, the virtual machine
directly inherits all of the features of that checkpoint-restart
package.  A further key difference of the new approach is that the
checkpoint-restart package operates {\em externally} to the virtual
machine.  Because it is not embedded inside the guest virtual machine
or the hypervisor, there is greater flexibility for it to interact as
a standard process within the hypervisor or host operating system.

As an example, a desktop user can now gain very high reliability
by running within a virtual machine that is set to take a snapshot
every minute.  Section~\ref{sec:experiment} demonstrates that the
DMTCP features of forked checkpointing and mmap-based fast restart
enable a virtual machine snapshot in 0.2~seconds, when running with the
Btrfs filesystem.  That section also shows a run-time overhead that
is too small to be measured when running the nbench2 benchmark program.
Btrfs is expected to become the default filesystem for Fedora, Ubuntu,
and others in about a year.

The generic mechanism of this work is based on the DMTCP checkpoint-restart
package~\cite{dmtcp09}.  The mechanism is demonstrated on three types
of virtual machines: KVM/QEMU~\cite{kvm}, user-space (standalone)
QEMU~\cite{qemu}, and Lguest~\cite{lguest}.  In all three cases, the
hypervisor (VMM --- virtual machine monitor) is based on Linux as the host
operating system.  The three examples cover three distinct situations:
entirely user-space virtualization (QEMU), full virtualization using
a Linux kernel driver (KVM/QEMU), and paravirtualization using a Linux
kernel driver (Lguest).  (A paravirtualized virtual machine is a virtual
machine that requires modifications to the host operating system.)

By providing checkpoint-restart capability to existing virtual machines,
one can retroactively add snapshot capability to a virtual machine in a
mostly transparent manner.  DMTCP already checkpoints user-space QEMU
``out of the box''.  An additional DMTCP-based plugin is required for
KVM/QEMU, but neither KVM/QEMU nor DMTCP is modified.  In the case of
Lguest, the kernel driver from Lguest require about 40~lines of new code
to support the checkpoint-restart capability.

The additional code required to checkpoint a new virtual machine is
approximately 200~lines (for plugins in the case of KVM/QEMU and Lguest).
Since user-space QEMU has no kernel driver component, DMTCP is able to
directly checkpoint it.

Given our experience it is estimated that someone familiar with the
examples provided here, could implement checkpoint-restart for a new
virtual machine in approximately five person-days --- assuming that the
VM provides a kernel driver API, as is the case for KVM.  (KVM is the
kernel driver component of a KVM/QEMU virtual machine.)  Where no kernel
driver API is provided, the development time is estimated at ten days,
due to the need to understand VM kernel driver internals, and augment
the existing API between driver kernel space and user space.

The two virtual machines above (KVM/QEMU and Lguest) require the estimated
effort primarily due to the need to save state within the kernel driver,
and then to appropriately restore and patch the state within the kernel
driver at the time of restart.

Surprisingly, DMTCP was able to checkpoint user-space QEMU directly,
with no requirements for new code or new plugins.  In hindsight, this is
attributed to the fact that user-space QEMU has no kernel driver, and so
no communication between kernel space and user-space.  In experiments,
DMTCP and QEMU were used to checkpoint both the Linux and Windows guest
operating systems ``out of the box'', with no additional modifications.

For all three virtual machines, DMTCP~\cite{dmtcp09,dmtcpSourceforge} is
used for purposes of checkpointing.  DMTCP is a widely used user-space
transparent checkpoint-restart package.  DMTCP was chosen in part for
the sake of its support for third-party plugins (see Section~\ref{sec:plugin}).

Of the three virtual machines on which generic checkpoint-restart is
demonstrated, to the best of our knowledge Lguest has not previously
been checkpointed.  QEMU provides a ``savevm'' command for directly
checkpointing.  KVM/QEMU has been previously checkpointed by modifying
existing features within KVM~\cite{ncaKVM11,hapcwKVM08} and making use
of the QEMU savevm command.

By checkpointing using an external, generic checkpoint-restart package,
one immediately inherits DMTCP's ability to take a consistent distributed
snapshot.  This is useful for analyzing any type of distributed computation.
Furthermore, there is the opportunity to {\em easily} extend this work in such
future directions as:
\begin{itemize}
 \item  fork-based checkpointing (quiesce the VM process,
and fork a child VM process to be checkpointed, while the parent
continues to execute, using the copy-on-write semantics of fork);
 \item heterogeneous checkpointing (checkpointing different virtual machines
and ``bare'' processes running a distributed computation); and
 \item incremental and differential checkpointing (checkpointing
only that part of RAM that has changed since the last checkpoint).
\end{itemize}
DMTCP has already been used for fault-tolerant applications,
while demonstrating each of the above features.  Thus, checkpoint-restart
of virtual machines can be extended to take advantage of such features.
At the same time, the checkpoint-restart capability remains largely
orthogonal to the ongoing internal development of the virtual machine
packages.

\paragraph{Snapshots (including filesystem).}
A distinction is sometimes made between checkpoints and snapshots when
the terminology is applied to virtual machines.  A {\em checkpoint} is
a copy of the state of a virtual machine suitable for being restored.
Such a checkpoint may or may not include saving a copy of the filesystem.
A {\em snapshot} always includes a full copy of the filesystem.

In a snapshot, rather than copy the entire filesystem during each
checkpoint, one prefers to use a filesystem supporting copy-on-write
in order to take a snapshot.  Filesystems that support copy-on-write
usually also support incremental snapshots.

This is a stable filesystem
that is likely to be readily available in most future Linux distributions.
Btrfs has been in the mainline Linux kernel since~2009 (since Linx~2.6.29).
Both Fedora and Ubuntu are planning for Btrfs to be the default
filesystem in late 2013 or later.

The experimental section uses copy-on-write incremental snapshots,
based on Btrfs~\cite{btrfs12}, for most experiments.  The time to take
a snapshot of the guest filesystem tends to be too small to measure as
part of the total restart time.

\paragraph{Forked checkpointing and fast restart:}
The checkpoint and restart can be sped up through standard features of
DMTCP.  At checkpoint time, forked checkpointing is employed.  The guest
VM (viewed as a process in the host) forks itself, and the child process
is checkpointed.  Fast restart uses mmap to map the checkpoint image
into RAM.  This allows the memory pages to be demand paged in as needed.
Forked checkpointing reduces the delay for a checkpoint to approximately
0.2~seconds (while the child process continues to write out the checkpoint
image), and the fast restart time is about 0.1~seconds.

In the rest of this paper, Section~\ref{sec:plugin} provides
background on DMTCP plugins.  Section~\ref{sec:ckptRestart} describes
the generic mechanism for checkpoint-restart of virtual machines.
Section~\ref{sec:implementation} describes several challenges in the
implementation, in order to provide deeper insights into the issues
in implementing the generic mechanism.   Finally, Section~\ref{sec:experiment} provides
experimental results, Section~\ref{sec:relatedWork} describes related
work, and Section~\ref{sec:conclusion} provides the conclusion.

\section{DMTCP Plugins:  Background}
\label{sec:plugin}

DMTCP (Distributed MultiThreaded CheckPointing)~\cite{dmtcp09} is used to
checkpoint and restart a virtual machine.  The current version of DMTCP
(DMTCP-1.2.6)~\cite{dmtcpSourceforge} provides a facility for third-party
plugins.  The work described here was based on DMTCP svn revision~1755.

When a new virtual machine is launched (\hbox{e.g.} QEMU), the user
prefixes the launch command with {\tt dmtcp\_checkpoint}.  A checkpoint
image is then created, and the virtual machine is restarted
via {\tt dmtcp\_restart}:
\begin{quotation}
    \noindent
    {\tt dmtcp\_checkpoint --with-plugin $\backslash$ \hbox{\ \ \ \ }
	dmtcp\_VM\_plugin.so qemu ...} \\
    {\tt dmtcp\_command --checkpoint} \\
    {\tt dmtcp\_restart qemu\_*.dmtcp} \\
\end{quotation}
\noindent
In the above scenario, VM would be KVM or LGUEST.  The plugin
{\tt dmtcp\_VM\_plugin.so} is the additional code developed for this work.

Plugins allow the functionality of DMTCP to be extended without modification
to the underlying DMTCP binary.
For the purposes of this work, we use two essential features of DMTCP plugins.
\begin{enumerate}
  \item {\em Wrapper functions:}  DMTCP provides wrapper functions around
	calls to library functions.
	In particular, it supports wrappers around system calls.
  \item {\em DMTCP event handling:}  DMTCP notifies the plugins of several
	events.  DMTCP blocks while plugins process events.
	The most important events for our purposes are
	pre-checkpoint and post-restart.
\end{enumerate}

A {\em wrapper function} is a function that is interposed between the
caller and a callee.  If the base code calls a function {\em foo}, and
if a wrapper function {\em bar} is interposed between the base code and
foo, then the base code calls bar instead.  DMTCP provides a mechanism
for plugins to transparently insert such wrapper functions around any
library call, including system calls.  In typical usage, the wrapper
function will then call the interposed functions (although possibly with
modified arguments), and then pass back a (possibly modified) copy of
the return value of the interposed function.  For a review of the many
techniques for interposition, see~\cite{ThainLivnyInterposition01}.

\begin{figure}[htb]
\begin{center}
\includegraphics[scale=0.8]{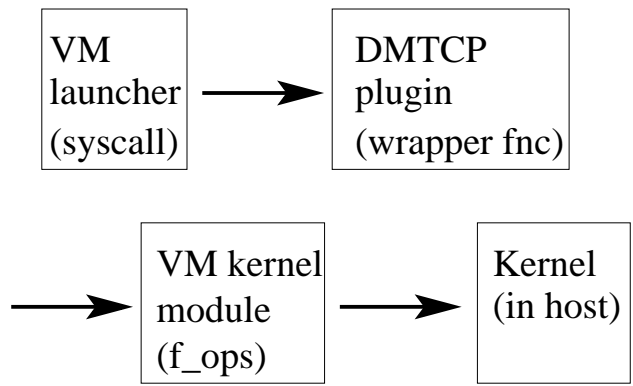}
\end{center}
\caption{A system call is initiated by a virtual machine launcher as it passes
 through DMTCP, the VM kernel driver in the host O/S, and then the kernel
 of the host O/S.  The DMTCP wrapper function allows DMTCP to record
 configuration information from the initial launch of the virtual machine,
 and then restore the original configuration at the time of restart.}
\label{fig:dmtcp-plugin}
\end{figure}

The VM kernel driver may interpose its own wrapper functions around
system calls that refer to a device supported by the virtual machine.
(Traditional kernel terminology does not call this a wrapper function.)
For example, KVM creates wrappers for the device {\tt /dev/kvm}, and
Lguest creates wrappers for the device {\tt /dev/lguest}.  Thus, the
DMTCP plugin is effectively creating its own wrapper function around
a VM-supplied wrapper, which in turn delegates to the kernel for the
standard functionality.

Next, we discuss DMTCP events.
During a pre-checkpoint event, all user threads have been quiesced,
and DMTCP has not yet begun to save the process state (including the
state of memory).  During a post-restart event, DMTCP has finished
restoring process state (including all of memory), but control has
not yet been returned to the user threads.

This design allows the plugin to save additional state relevant to the
virtual machine at the time of checkpoint.  During the post-restart event,
the checkpoint-restart appears transparent to the plugin.  Hence, the
plugin finds the VM state in whatever data structure that the plugin
had originally used to save the information.

The plugin uses a virtual-machine-specific method to transfer data between
the kernel driver of the virtual machine and the user-space memory where
the plugin ``lives''.  Sections~\ref{sec:API-KVM} and~\ref{sec:API-Lguest}
describes those VM-specific mechanisms.

A typical use of a DMTCP plugin is to use a wrapper function
to record information by certain system
calls issued by the launcher.  This allows the DMTCP plugin to execute
modified versions of those same system calls at the time of restart,
before the thread of control is handed back to the virtual machine.

\begin{figure*}[htb]
\begin{center}
\includegraphics[scale=0.8]{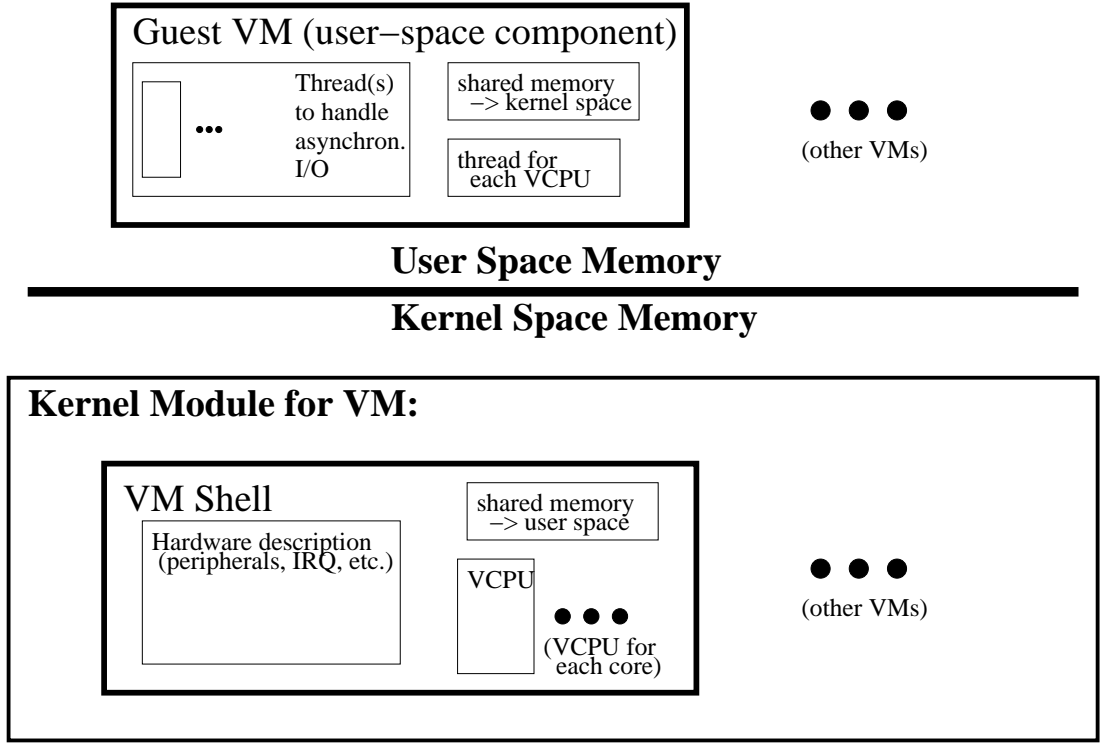}
\end{center}
\caption{Launching a Virtual Machine:  a Generic Architecture.  This sketch
  illustrates the components of interest for checkpoint-restart.
  The {\em VM shell} refers to one or more uninitialized data structures in the
  kernel driver that describe the virtual machine.  A VM launcher will
  initialize those data structures, and a generic checkpoint-restart mechanism
  must be prepared to restore those data structures appropriately.
  }
\label{fig:vm-launch}
\end{figure*}

\section{Generic Mechanism for Checkpoint-Restart}
\label{sec:ckptRestart}

In this section, we describe the general mechanism for checkpointing
and restarting a virtual machine.  In the rest of this section,
Section~\ref{sec:pluginOverview} provides an overview of the
actions of our DMTCP plugin in supporting checkpoint-restart.
Section~\ref{sec:launch} describes a generic sequence of steps that
any virtual machine must employ in launching a new virtual machine.
Section~\ref{sec:relaunch} then describes the steps needed to restore and
restart that virtual machine.  Finally, the generic mechanism depends
on the APIs provided by the virtual machine (or augmented APIs in the
case of Lguest).  Sections~\ref{sec:API-KVM} and~\ref{sec:API-Lguest}
describe those APIs for KVM and Lguest, respectively.  The APIs are
responsible for saving the VM driver state, and later for launching a
shell VM, and then restoring the VM driver state.

The existing DMTCP package already transparently checkpoints and
restores all of user-space memory, along with essentially all pertinent
process state (threads, open file descriptors, associated terminal
device, stdin/stdout/stderr, sockets, shared memory regions, etc.).
Where a subsystem refers to an external object, DMTCP has several
subsystem-specific heuristics for restoring such information.  Examples
of such cases abound:  open files that were modified or re-named after
checkpoint; sockets to database servers; shared memory regions with
daemons such as NSCD; etc.

In the case of user-space QEMU, the existing DMTCP package and its
heuristics for restoring subsystems sufficed to correctly checkpoint
and restart QEMU.  No DMTCP plugin was required.
This was tested with QEMU running each of Linux
and Microsoft Windows.  (See Section~\ref{sec:experiment}.)
The rest of this section is concerned with KVM/QEMU and Lguest,
for which a DMTCP plugin was required.

\subsection{Overview of Actions of DMTCP Plugin}
\label{sec:pluginOverview}

For the KVM/QEMU and Lguest virtual machines, a DMTCP plugin was
implemented to save and restore state contained in the VM kernel
driver.  Recall from Section~\ref{sec:plugin} that the two features
of DMTCP plugins we use are wrapper functions and notification of the
pre-checkpoint/post-restart events.  Wrapper functions are used to record
information about system calls sent by the VM launcher to the kernel.
At the time of pre-checkpoint, the plugin saves certain state within the
VM kernel driver.  Since that information is contained in the plugin's
user-space memory, DMTCP automatically saves it at checkpoint time and
later restores it, as part of DMTCP's standard procedure for saving and
restoring all of user-space memory.  At the time of post-restart, the
plugin copies that state back into the VM kernel driver and appropriately
patches it.

In addition to implementing a VM-specific plugin, one must modify about
40~lines in in Lguest (in {\tt lguest\_user.c}).  This is because KVM
provides an API for communication between the VM kernel driver and
the DMTCP plugin.  Lguest does not.  Hence, we have augmented the API
of Lguest.  This is needed to enable the plugin to save and restore state.

\bigskip
In overview, the DMTCP plugin is needed only in the case of VM kernel
drivers (KVM/QEMU and Lguest in this work).  It does the following:
\begin{enumerate}
  \item {\em Time of Original VM Launch:}\/ Wrapper functions in the DMTCP
	plugin record pertinent information from the system calls made by
	the VM launcher.  This information is used to later restore
	the configuration of memory, etc., of the new virtual machine
	created by the VM launcher.
  \item {\em Checkpoint Time:}\/ The DMTCP plugin is notified of the
	pre-checkpoint event after the user threads have been quiesced,
	and before all of user-space memory is copied to a checkpoint image.
	The DMTCP plugin then copies pertinent information from the
	data structures inside the VM kernel driver.  This uses a kernel
	driver API to user space (KVM/QEMU), or else an augmented driver
	provided by us (Lguest).
  \item {\em Restart Time (restoring user-space memory of the VM):}\/
	DMTCP restores user-space memory to the same addresses where they
	existed prior to checkpoint.  DMTCP does this transparently,
	and the DMTCP plugin does not do any work at this stage.
  \item {\em Restart Time (re-launching the VM):}\/ The DMTCP plugin
	is notified when all user-space memory and process state has been
	restored, but before control is returned to the user threads.
	At this time, the user-space component of the VM has been fully
	restored.  But the pre-checkpoint VM does not exist, and so
	the VM kernel driver is not aware of any VM's.  The DMTCP plugin
	replays a modified version of the first few system calls by the
	VM launcher.  It replays just enough to provide an ``empty shell''
	of a virtual machine.  Many of the VM kernel driver data structures
	have not been initialized, and for some data structures, not even
	storage has been allocated.
  \item {\em Restart Time (patching the VM kernel driver):}\/ The DMTCP plugin
	must now copy its saved VM kernel driver state back into the
	VM kernel driver.  However, in some cases, that VM kernel driver
	state must be modified to account for the fact that this is not
	the original VM, and the kernel may have changed some of the
	memory addresses in this re-launched VM.  This uses a kernel
	driver API to user space (KVM/QEMU), or else an augmented driver
	provided by us (Lguest).
\end{enumerate}

\subsection{Launching a Virtual Machine}
\label{sec:launch}

We first describe in general terms how a virtual machine is launched
(created).  Any particular virtual machine may differ in some detail,
or may merge or sub-divide the steps described.  It is partly for
this reason, that we do not currently see the possibility of a fully
transparent checkpoint-restart mechanism.  However, this general
description provides a framework that can be used to accelerate the
development of a plugin for a new virtual machine.

Figure~\ref{fig:vm-launch} shows those portions of a virtual machine
of interest for checkpoint-restart.  Typically, the virtual machine is
created by a command issued from user-space.  The program run by that
command is referred to as a {\em VM launcher}, which sets up, runs and
services the Guest.

The launcher must:
\begin{enumerate}
	\item open an interface to the host kernel via a character
	device (\hbox{e.g.} {\tt /dev/kvm} or {\tt /dev/lguest}).
	\item initialize the VM:  tell the kernel where is the start of
	 the guest physical memory in the launcher's virtual address
	space.
	\item arrange to virtualize IRQ interrupts.
	\item create and initialize virtual CPUs to hold the current state
	of the registers.
	\item run the guest.
\end{enumerate}

\subsection{Re-Starting a Virtual Machine from a Checkpoint Image}
\label{sec:relaunch}

\begin{figure*}[htb]
\begin{center}
\includegraphics[scale=0.8]{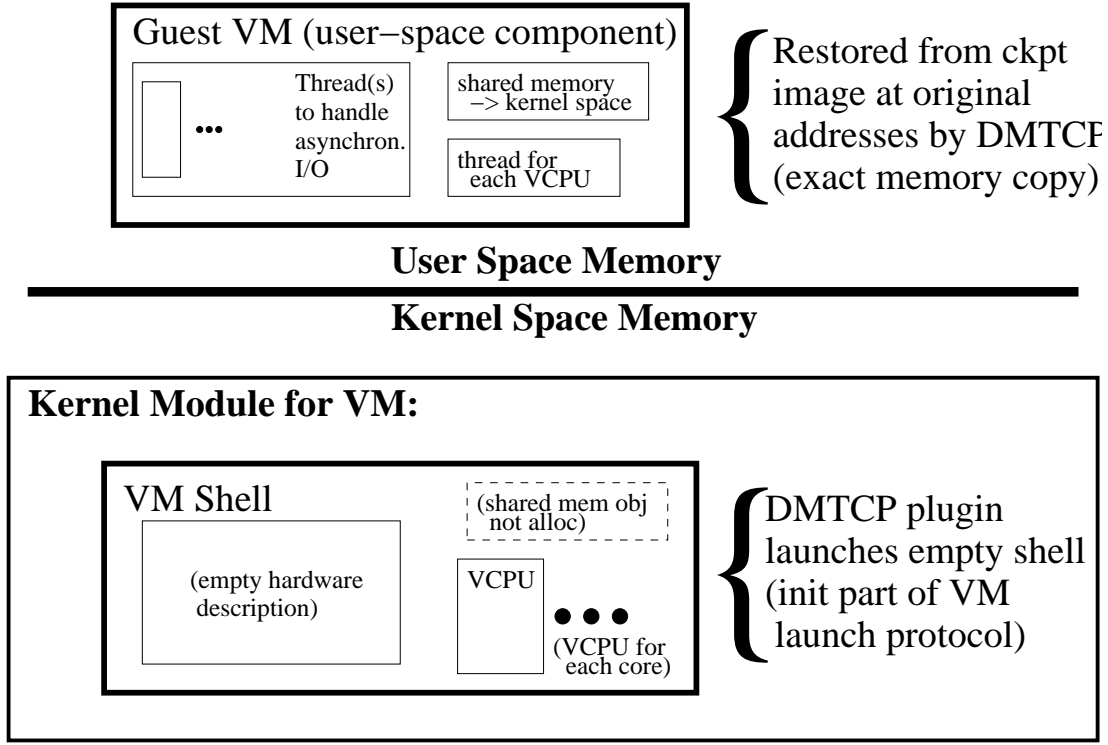}
\end{center}
\caption{Re-Starting Virtual Machine from Checkpoint Image.  The
 original hardware description in the kernel driver must be re-created.
 In addition, the mechanism for sharing memory between the kernel
 and the user-space VM component must be re-created.}
\label{fig:vm-restart}
\end{figure*}

In restoring the memory after checkpoint, the user-space memory (memory
of QEMU) is restored exactly.  The memory within the VM kernel driver
must be restored in one of three ways.
\begin{enumerate}
  \item {\em Launch a ``shell'' of a new VM in kernel driver:}\/
	On restart, the DMTCP plugin executes the first few steps of
	launching a VM, in order to create an empty shell for the
	VM data structures.  (See Figure~\ref{fig:vm-restart}.)
	We refer to this as ``re-launch''.
  \item {\em Restore pre-checkpoint state of kernel driver:}\/
	Next, we identify those data structures of the VM kernel driver
	that have not yet been initialized.  For those data structures,
	we design the DMTCP plugin to save the values at the time
	of checkpoint, and restore the values at the time of restart.
  \item {\em Patch kernel driver state:}\/
	Finally, some of the data values that are restored are incorrect.
	These must be filled in correctly on a case-by-case basis.
	For example, at the time of launching a new VM, KVM dynamically
	allocates memory for a struct that describes the memory addresses
	where user-space QEMU resides.  At the time of restart, the
	kernel is unlikely to allocate the new struct at the same
	address as before.  The DMTCP plugin must save the data
	from the old struct prior to checkpoint and use it within
	the new struct allocated at the time of restart.  KVM provides
	an API for this purpose, while for Lguest the API was augmented.
\end{enumerate}

Figure~\ref{fig:vm-restart}

\subsection{Case History:  APIs used by DMTCP Plugin for KVM}
\label{sec:API-KVM}
	
QEMU uses KVM's ioctl commands to check for the 
different hardware capabilities and configures data structures 
internal to the kernel driver. They represent the state of the 
virtual machine. Once configured, these data structures can be read 
from QEMU using ioctl system calls with different {\tt GET\_XXX} parameters.
The DMTCP plugin retrieves values of 
relevant data structures for task state segment address,  guest 
registers, a programmable interval timer, the IRQ chip and the registers
of the virtual CPU.

For certain internal kernel driver data structures, there is a {\tt SET\_XXX}
parameter, but no
corresponding {\tt GET\_XXX} parameter.  Hence, the DMTCP plugin
defines a wrapper function around ioctl, and monitors the initialization
of the missing data structures via calls by the VM launcher of ioctl.
Upon restart, the plugin (running inside the launcher process) issues
an ioctl call with
the appropriate {\tt SET\_XXX} parameter and the appropriate values discovered
during the original launch.

\subsection{Case History:  APIs used by DMTCP Plugin for Lguest}
\label{sec:API-Lguest}

Lguest already employs the read and write system calls to pass the
parameters needed in VM launch as described in Section~\ref{sec:launch}.
These system calls were extended to provide an API for reading and writing
internal data structures of the kernel driver.  Another alternative would
have been to augment the API using ioctl, similarly to the situation
under KVM.  Some of the data structures saved and restored are the
virtual cpu, registered eventfd objects,and the address of the guest
physical address, stack, page directory, etc.

\section{Implementation Challenges}
\label{sec:implementation}

Two particular implementation issues required special treatment.

\paragraph{Graphics support}
  In checkpoint-restart of virtual machines, most modern operating systems
  support GUI interfaces.  Hence, the graphics of the GUI must be
  checkpointed and restored.  A standard trick is used.  The virtual
  machine is run inside TightVNC, an example of a VNC client-server for
  virtual network computing. In particular, QEMU starts up a vncserver
  for the graphics at the time of launching.  A VNC viewer connects to the
  VNC server.  Just prior to checkpoint, we disconnect the VNC viewer, and
  we reconnect after resuming or restarting the guest VM. 

\paragraph{Anonymous inodes for KVM}
    When KVM launches a new virtual machine, it maps a region from kernel space
to user-space memory for convenience of communication between the kernel-level
KVM driver and the user-space QEMU component.  This is implemented by having
QEMU call mmap on an anonymous inode.  (An anonymous inode will be deleted when
no object continues to refer to it.  Since the anonymous inode is associated with
the KVM node, when QEMU calls mmap, the call is intercepted by the KVM driver,
and it arranges for the kernel space that will be mapped into user space.)
This occurs when KVM launches a new
virtual machine.

At the time of restart, the DMTCP plugin must then re-create the memory
region for sharing between user space and kernel space.  
Since the user-space component (QEMU) will be restored exactly at restart
time, it will retain pointers into the address of the
mapped region backed by the anonymous inode, as it existed prior to checkpoint.

The DMTCP plugin handles this issue in two phases:  prior to checkpoint,
and at the time of restart.  Prior to checkpoint (during the original
launch), the wrapper for mmap inside the DMTCP plugin detects the call
by QEMU for the specific anonymous inode in question.  The return value
of mmap is then saved by the DMTCP plugin.  At the time of restart,
the DMTCP plugin calls mmap, and specifies the anonymous inode and the
original address where it had been mapped.  In addition, the DMTCP plugin
mmap wrapper invokes the parameter {\tt MAP\_FIXED} in order to re-map
the region at the desired address.

\begin{table*}[htp]
\centering
  \scriptsize
  \addtolength{\tabcolsep}{-3pt}
  \begin{tabular}{|c|c|c|c|c|c|c|c|c|c|c|}
   \hline
   \multicolumn{1}{|c|}{Allocated} & \multicolumn{1}{|c|}{Free} & \multicolumn{3}{|c|}{Lguest} & \multicolumn{3}{|c|}{KVM/QEMU} & \multicolumn{3}{|c|}{QEMU (user-space)}\\
   \cline{3-11}
    Mem. (MB) & Mem. (MB) & Ckpt (s) & Restart (s) & Image
      & Ckpt (s) & Restart (s) & Image
      & Ckpt (s) & Restart (s) & Image\\
   \hline
    128 & 2.5 & 2.292 & 1.264 & 30 MB & 3.949 & 1.308 & 44 MB & 4.342 & 1.686 & 59 MB \\
    \hline
    256 & 4.2 & 3.169 & 1.382 & 33 MB & 6.424 & 2.353 & 89 MB & 7.705 & 3.017 & 109 MB \\
    \hline
    512 & 184 & 5.390 & 2.417 & 35 MB & 9.886 & 3.278 & 129 MB & 11.870 & 4.427 & 170 MB \\
    \hline
    768 & 441 & 6.823 & 3.013 & 38 MB & 9.212 & 3.307 & 130 MB & 14.039 & 5.047 & 194 MB \\
    \hline
    1024 & 700 & 8.339 & 2.986 & 37 MB & 10.033 & 3.130 & 122 MB & 16.504 & 5.467 & 208 MB\\
    \hline
  \end{tabular}
\caption{\label{tab:ckpt} Checkpoint-restart times for idle virtual machines.
    The checkpoint times include the times for compressing the memory image and
    writing the contents to the disk.}
\end{table*}

\begin{table*}[htp]
\centering
  \scriptsize
  \addtolength{\tabcolsep}{-3pt}
  \begin{tabular}{|c|c|c|c|c|c|c|c|c|c|}
   \hline
   \multicolumn{1}{|c|}{Allocated} & \multicolumn{3}{|c|}{Lguest} & \multicolumn{3}{|c|}{KVM/QEMU} & \multicolumn{3}{|c|}{QEMU (user-space)}\\
   \cline{2-10}
      Memory (MB) & Ckpt (s) & Restart (s) & Image Size
      & Ckpt (s) & Restart (s) & Image Size
      & Ckpt (s) & Restart (s) & Image Size\\
   \hline
    128 & 0.157 & 1.183 & 30 MB & 0.183 & 1.284 & 44 MB & 0.161 & 1.697 & 59 MB \\
    \hline                    
    256 & 0.174 & 1.426 & 32 MB & 0.200 & 2.379 & 90 MB & 0.165 & 2.985 & 111 MB \\
    \hline                    
    512 & 0.176 & 2.523 & 35 MB & 0.233 & 3.061 & 122 MB & 0.174 & 4.435 & 171 MB \\
    \hline                    
    768 & 0.174 & 2.447 & 36 MB & 0.211 & 3.106 & 122 MB & 0.183 & 4.970 & 191 MB \\
    \hline                    
    1024 & 0.178 & 2.818 & 37 MB& 0.243 & 2.964 & 116 MB & 0.191 & 5.633 & 213 MB\\
    \hline
  \end{tabular}
\caption{\label{tab:fork-ckpt} Forked checkpoint-restart times
    for idle virtual machines.
    (The size of free memory is the same as for Table~\ref{tab:ckpt}.)}
\end{table*}

\section{Experimental Results}
\label{sec:experiment}

\subsection{Configuration}
We ran our experiments on a system with an Intel Core~i7 (2.3~GHz)
and 8~GB of RAM.  This was part of a MacBook laptop with a 256~GB SSD.
The host operating system was a 32-bit version of Ubuntu-12.10 with Linux
kernel-3.5.7. The host was running natively in its own partition on
the MacBook. The guest was set up to run Ubuntu-8.04. DMTCP svn revision 1755 was used for all experiments.

All experiments represent full snapshots, including a snapshot of
the guest filesystem. The guest filesystem appears as a single
file within the host filesystem. Unless otherwise noted, the guest filesystem
is located within a Btrfs filesystem of the host operating system.
Checkpoint includes the time to create a snapshot of the guest
filesystem within Btrfs.  The snapshot of the guest filesystem is created
using the GNU binutils command ``{\tt cp --reflink}''.  This operation
tends to be fast, since it primarily involves taking a snapshot of
the current data blocks of the host file comprising the guest filesystem.

Experiments were conducted for:  broad coverage
(Section~\ref{sec:coverage}); forked checkpointing
(Section~\ref{sec:forkedCkpt}); tests of DMTCP features
for forked checkpointing and mmap-based fast restart
(Section~\ref{sec:fastRestart}); analyzing the impact of running the
nbench2 benchmark program (Section~\ref{sec:benchmarks}); and the overhead of
saving snapshots of the guest filesystem on a host Btrfs filesystem
(Section~\ref{sec:btrfs}).

\subsection{Coverage tests}
\label{sec:coverage}

Table~\ref{tab:ckpt} demonstrates the memory-intensive version
of checkpoint-restart using the default mode of DMTCP (using gzip
compression) on an idle virtual machine.  The checkpoint times grow
roughly proportionally to the size of the allocated memory for the larger
sizes (512~MB guest~VM to 1024~MB guest~VM).  Below those memory sizes,
other factors in the checkpoint times presumably dominate. Restart times
do not change appreciably at the higher ranges of memory.

\subsection{Forked checkpointing}
\label{sec:forkedCkpt}

Forked checkpointing on an idle virtual machine is
demonstrated in Table~\ref{tab:fork-ckpt}.  This uses the ``{\tt
--enable-forked-checkpointing}'' configure option of DMTCP, such that
at checkpoint time, a child process is created. The child fulfills the
rest of the checkpoint, while the parent process continues computing
concurrently.  As would be expected, the parent completes its portion
of the checkpoint largely independently of the size of the checkpoint
image or allocated memory. Forked checkpointing typically requires
about 0.2~seconds. Since the checkpoint was taken while the virtual
machine was running, it was not possible to take checkpoints at the
same time within the two runs (forked checkpointing and standard).
For this reason, the sizes of the images differ by approximately~2.5\%,
as seen in Table~\ref{tab:ckpt}.

The times for checkpoint and restart for KVM/QEMU are larger than the
times for user-space QEMU. This is because the plugin for KVM/QEMU makes
extra system calls at checkpoint and restart time.  The times can be
reduced by modifying the kernel driver to implement a new system call
that coalesces all of the operations of the previous system calls.

\subsection{Fast Restart}
\label{sec:fastRestart}

\begin{table*}[htp]
\centering
  \scriptsize{
  \addtolength{\tabcolsep}{-3pt}
  \begin{tabular}{|c|c|c|c|c|c|c|c|c|c|}
   \hline
   \multicolumn{1}{|c|}{Allocated} & \multicolumn{3}{|c|}{Lguest} & \multicolumn{3}{|c|}{KVM/QEMU} & \multicolumn{3}{|c|}{QEMU (user-space)}\\
   \cline{2-10}
      Memory (MB) & Ckpt (s) & Restart (s) & Image Size 
      & Ckpt (s) & Restart (s) & Image Size
      & Ckpt (s) & Restart (s) & Image Size\\
   \hline
    128 & 0.523 & 0.096 & 139 MB & 0.689 & 0.097 & 182 MB & 0.593 & 0.098 & 230 MB \\
    \hline
    256 & 0.834 & 0.098 & 267 MB & 1.098 & 0.092 & 311 MB & 1.329 & 0.096 & 408 MB \\
    \hline
    512 & 1.489 & 0.097 & 523 MB & 1.843 & 0.098 & 566 MB & 2.437 & 0.097 & 761 MB \\
    \hline
    768 & 2.495 & 0.097 & 779 MB & 2.523 & 0.094 & 823 MB & 3.539 & 0.096 & 1.1 GB \\
    \hline
    1024 & 3.021 & 0.098 & 1.1 GB & 3.119 & 0.098 & 1.1 GB & 4.480 & 0.097 & 1.5 GB \\
    \hline
  \end{tabular}
  }
\caption{\label{tab:fast-ckpt} Fast restart times for
 idle virtual machines.
    (The size of free memory is the same as for Table~\ref{tab:ckpt}.)}
\end{table*}

\begin{table*}[htp]%\small
\centering
  \scriptsize{
  \addtolength{\tabcolsep}{-3pt}
  \begin{tabular}{|c|c|c|c|}
   \hline
   \multicolumn{1}{|c|}{Allocated Memory} & \multicolumn{3}{|c|}{QEMU/KVM} \\
   \cline{2-4}
      (MB) & Checkpoint (s) & Restart (s) & Image Size \\
   \hline
    128 & 0.204 & 0.095 & 184 MB  \\
    \hline
    256 & 0.194 & 0.093 & 310 MB  \\
    \hline
    512 & 0.205 & 0.095 & 568 MB  \\
    \hline
    768 & 0.223 & 0.098 & 822 MB  \\
    \hline
    1024 & 0.206 & 0.095 & 1.1 GB \\
    \hline
  \end{tabular}
  }
\caption{\label{tab:fcr} Forked checkpoint and fast restart times for
 an idle VM under QEMU/KVM.
    (The size of free memory is the same as for Table~\ref{tab:ckpt}.)}
\end{table*}

Table~\ref{tab:fast-ckpt} employs fast restart on an idle virtual
machine using the ``{\tt --enable-fast-ckpt-restart}'' option of DMTCP.
This option uses mmap to map the checkpoint image directly into memory,
instead of copying it.  In this case, memory is demand-paged in as
needed from the checkpoint image. In this mode, compression is not used
in creating the checkpoint image. Checkpoint times are somewhat faster
in writing an uncompressed checkpoint image to disk, since the time for
executing gzip (compression) dominates over the time to write to disk.

Table~\ref{tab:fcr} presents the results of combining both fast-restart
and forked-checkpointing mechanisms on QEMU/KVM. Note that on restart
from a checkpoint image, the shadow page tables inside the kernel must be
recreated, after which the pages will be faulted back into RAM. The impact
of this on the performance of the running applications within the guest
operating system is not captured by these tables. The tables indicate
only the time after which the virtual machine can begin to execute.

\subsection{The nbench2 benchmark program}
\label{sec:benchmarks}

The numbers in Table~\ref{tab:nbench2} demonstrate the small overhead
of executing with DMTCP. DMTCP incurs this overhead due to its use
of wrapper functions around certain system calls. We used the nbench2
benchmark program~\cite{nbench} to analyze the overhead under conditions
of stress. The nbench2 benchmark program is a collection of applications
that stress the cpu and the memory. The applications stress the integer
unit, the floating-point unit and the memory subsystem. The indexes in
Table~\ref{tab:nbench2} are a measure of performance, normalized with
respect to the AMD~K6/233. Higher numbers are better.

Table~\ref{tab:nbench2} shows that DMTCP has little impact on performance
for a VM running cpu-intensive or memory-intensive loads. In contrast the
performance of KVM/QEMU is much higher than user-space QEMU, as expected.

Table~\ref{tab:benchmarks} shows the large impact of using DMTCP
optimizations to enhance the checkpoint and restart times. Further, one
can compare the effect of running a virtual machine under load with an
idle virtual machine. Table~\ref{tab:benchmarks} shows a machine under
load (running nbench2), while Tables~\ref{tab:ckpt}, \ref{tab:fork-ckpt}
and~\ref{tab:fast-ckpt} show an idle machine. The checkpoint and restart
times are almost the same in the two cases. The size of the checkpoint
image increases by at most~7.2\% when under load. This is due to fewer
zero page when under load.

\begin{table*}[htp]
\centering
  \scriptsize{
  \addtolength{\tabcolsep}{-3pt}
  \begin{tabular}{|c|c|c|c|c|c|c|}
   \hline
   \multicolumn{1}{|c|}{} & \multicolumn{3}{|c|}{KVM/QEMU} & \multicolumn{3}{|c|}{QEMU (user-space)}\\
   \cline{2-7}
      & Memory Index & Integer Index & Floating-point Index
      & Memory Index & Integer Index & Floating-point Index \\
   \hline
    With DMTCP & 31.483 & 25.535 & 47.806 & 2.516 & 3.473 & 0.285 \\
    \hline
    Without DMTCP & 31.381 & 25.518 & 48.380 & 2.435 & 3.338 & 0.274 \\
    \hline
  \end{tabular}
  }
\caption{\label{tab:nbench2} Nbench2 Benchmark program on Virtual Machines.
  (Memory allocated in each case is 1024~MB.)}
\end{table*}

\begin{table*}[htp]
\centering
  \scriptsize{
  \addtolength{\tabcolsep}{-3pt}
  \begin{tabular}{|c|c|c|c|c|c|c|}
   \hline
   \multicolumn{1}{|c|}{Checkpoint Mechanism} & \multicolumn{3}{|c|}{KVM/QEMU} & \multicolumn{3}{|c|}{QEMU (user-space)}\\
   \cline{2-7}
      & Checkpoint (s) & Restart (s) & Image Size
      & Checkpoint (s) & Restart (s) & Image Size\\
   \hline
    Default-ckpt & 9.915 & 3.203 & 125 MB & 15.154 & 5.967 & 226 MB \\
    \hline
    Forked-ckpt & 0.214 & 3.171 & 125 MB & 0.188 & 5.902 & 226 MB \\
    \hline
    Fast-restart & 3.245 & 0.098 & 1.1 GB & 4.382 & 0.093 & 1.5 GB \\
    \hline
    Forked-ckpt/Fast-restart & 0.206 & 0.095 & 1.1 GB & 0.212 & 0.122 & 1.5 GB \\
    \hline
  \end{tabular}
  }
\caption{\label{tab:benchmarks} Checkpoint-restart while running nbench2, as influenced by DMTCP options for forked checkpoint and fast restart.
  (Memory allocated in each case is 1024~MB.)}
\end{table*}

\begin{table*}[htp]%\small
\centering
  \scriptsize{
  \addtolength{\tabcolsep}{-3pt}
  \begin{tabular}{|c|c|c|}
   \hline
       & Checkpoint (s) & Restart (s)  \\
   \hline
    Btrfs & 0.264 & 0.102 \\
    \hline
    Without Btrfs & 7.932 & 8.428 \\
    \hline
  \end{tabular}
  }
\caption{\label{tab:btrfs} Snapshotting an idle guest VM under KVM/QEMU,
including its guest filesystem. The guest filesystem is optionally stored
in a host Btrfs filesystem.
   (Memory allocated in each case is 1024~MB. Size of guest filesystem
   is 2.5~GB.)}
\end{table*}

\subsection{Btrfs}
\label{sec:btrfs}

Table~\ref{tab:btrfs} shows the advantage of using the copy-on-write
feature of Btrfs to store the guest VM's filesystem. At checkpoint time a small
additional DMTCP plugin rapidly copies the state of the entire filesystem
(which appears as a single file on the host filesystem), using the {\tt
--reflink} option of the GNU binutils copy command. At restart time the
state of the guest filesytem is similarly copied back.

\section{Related Work}
\label{sec:relatedWork}

Checkpoint-restart mechanisms specific to individual virtual machines
have existed for some years. Xen~\cite{xen03} and QEMU~\cite{qemu}
are notable examples of this.

Xen has offered checkpointing at least since~\cite{ornlXen06}.
A faster checkpoint-restart based on COW (copy-on-write filesystems)
was developed independently by two groups~\cite{gatechXen10,ubcXen09}.
Later, support for deduplication in Xen checkpoints was described
in~\cite{veeXen11}.

QEMU can be checkpointed by issuing the ``stop'' command, followed by the
``savevm'' command.  This capability has been enhanced in the case of
QEMU running on top of KVM (kernel-based virtual machine).  This was done
by modifying KVM to add an additional checkpoint thread~\cite{ncaKVM11}
and similarly by modifying the live migration facility of KVM to save
a copy while migrating the ongoing VM computation~\cite{hapcwKVM08}.

For the support of snapshots, one requires a copy-on-write filesystem.
A common current choice is QCOW2~\cite{qcow2}, which supports
the creation of incremental snapshots.  Another recent choice is
BlobSeer~\cite{BlobSeer11}, as used in~\cite[Section~3.3]{BlobCR11}.
That choice has the advantage of exposing the raw checkpoint image file
to the host operating system or hypervisor.

The work described here uses Btrfs~\cite{btrfs12}.  Like BlobSeer, Btrfs
exposes the raw checkpoint image to the host, making it compatible with
the use of DMTCP from outside both the VM and the VM kernel driver.
Btrfs is a mainstream filesystem (in the mainline Linux kernel since
Linx~2.6.29).  Both Fedora and Ubuntu are planning for Btrfs to be the
default filesystem in late 2013 or later.

An alternative to using a copy-on-write filesystem for snapshots is the
use of a stackable filesystem.  This was discussed in~\cite{ornlXen06},
with the idea of using UnionFS~\cite{unionfs06}.  They appear not to
have implemented it.

DMTCP~\cite{dmtcp09} was chosen for checkpoint-restart due to its
recent support for plugins.  This eased the job of checkpoint-restart,
since other choices would have required modification to the underlying
checkpoint-restart package.  In addition, DMTCP's support for forked
checkpointing, and for fast restart (based on mmap), were also helpful
in demonstrating those features for virtual machine snapshots.

Among other choices for checkpoint-restart, BLCR~\cite{BLCR06}
has the longest history among the commonly used checkpoint-restart
packages.  It is based on a kernel module, and has especially strong
support for use with MPI-based checkpoint-restart services and with
batch queues.  CryoPid2~\cite{cryopid2} represents an alternative
user-space checkpoint-restart package based on using ptrace to control
the target application.  OpenVZ~\cite{openvz} is a kernel-based
checkpoint-restart package based on Linux containers.  CRIU~\cite{criu}
is a recent checkpoint-restart package with an interesting hybrid
strategy between user-space and kernel-space approaches.  The Linux
kernel has been extended to include many interfaces that expose the
kernel internals.  CRIU uses those interfaces to provide an entirely
user-space checkpoint-restart package.

Forked checkpointing has an exceptionally long history, dating
back to~1990~\cite{ForkedCheckpointing90,ForkedCheckpointing94}.
Incremental checkpointing has been demonstrated at least
since 1995~\cite{IncrementalCheckpoint95}.

\section{Conclusion}
\label{sec:conclusion}

A generic checkpoint-restart mechanism was presented based on the
DMTCP checkpoint-restart package.  DMTCP can directly checkpoint the
user-space QEMU virtual machine.  In other cases, where the virtual
machine employs a kernel driver, DMTCP relies on an API to transfer
driver state between the kernel driver and user space.  KVM provides
such an API, and so a 200-line DMTCP plugin sufficed to implement
checkpoint-restart for KVM/QEMU.  Lguest does not provide such an API,
and about 40~lines were added to augment the Lguest kernel driver.
The estimated development time for developing checkpoint-restart
for a new virtual machine is estimated at five person days (where
a full kernel driver API is provided, as for KVM), and ten person
days (where a full kernel driver API is not provided, as for Lguest).

The method is applicable wherever DMTCP is available.  DMTCP currently
runs under Linux (x86, x86\_64, and ARM).  Thin hypervisors may or may
not support DMTCP, depending on what features of Linux they support.

The generic mechanism presented assumes a homogeneous architecture
(same CPU, same host operating system, same hardware).  Future work
may consider removing some of those restrictions --- especially those
of homogeneous hardware.  Future work will also explore transparently
checkpointing a cluster of virtual machines.

Where the kernel driver API must be extended (Lguest, in our case),
an alternative approach was considered.  In this approach, {\em all
of} the system calls from the VM launcher to the VM kernel driver
are recorded at the time of launch, and more of those system calls are
played back at the time of restart (although possibly in modified form).
This may have advantages in being more robust as the VM software evolves.
This also is a topic for future work.

\section*{Acknowledgment}
The authors gratefully acknowledge the discussions and insights
  provided by Zhengping Jing.

\bibliographystyle{alpha}

\bibliography{dsn-ckpt-vm}

\end{document}